\documentclass[11pt]{article}
\usepackage{graphics,epsfig}
\usepackage{graphicx}
\usepackage{epstopdf}
\usepackage{amssymb}
\usepackage{amsmath}
\begin{document}

\title{Necessity of Combining
Mutually Incompatible Perspectives
in the Construction of a Global View:
Quantum Probability and Signal Analysis}
\author{Sven Aerts and Diederik Aerts \\
        \normalsize\itshape
        Center Leo Apostel for Interdisciplinary Studies (CLEA) \\
         \normalsize\itshape
         and Foundations of the Exact Sciences (FUND) \\
		\normalsize\itshape
		Department of Mathematics, Vrije Universiteit Brussel \\
        \normalsize\itshape
         1160 Brussels, Belgium \\
        \normalsize
        E-Mails: \textsf{saerts@vub.ac.be, diraerts@vub.ac.be}\\ \\ 
		Franklin E. Schroeck\\
		\normalsize\itshape
		Department of Mathematics \\
		\normalsize\itshape
		John Greene Hall, 2360 S. Gaylord St.\\
		\normalsize\itshape
		University of Denver, Denver, USA\\
		\normalsize
		E-Mail: \textsf{fschroec@du.edu}
		}
\date{}
\maketitle

\begin{abstract}
\noindent
The scientific fields of quantum mechanics and signal-analysis originated within different settings, aimed at different goals and started from different scientific paradigms. Yet the development of the  two subjects has become increasingly intertwined. We argue that these similarities are rooted in the fact that both fields of scientific inquiry had to  deal with finding a single description for a phenomenon that yields complete information about itself only when 
we consider mutually incompatible accounts of that phenomenon.
\end{abstract}

\noindent
Keywords: quantum probability, signal analysis, incompatible observables, observation, global view.

\section{Introduction}
Can we find a single common framework for all knowledge that in some way
converges to the truth, or should we perhaps accept that truth is inevitably
only meaningful with respect to the one who pronounces it, to the
context in which it is produced, to the frame of reference it is presented
in? The simple question we pose here is well-known to lead to endless
discussion on the nature of reality and knowledge, the relation between
ontology and epistemology and between theory and fact. This discussion
itself is an illustration of the dichotomic status of truth. The various
stances taken by the participants in the discussion contribute to the idea
of a relative truth, but the mere fact that we are still discussing the
issue, highlights that we still have not given up on trying to convince each
other that one truth is perhaps \textquotedblleft more\textquotedblright\
true, than another. In the face of all various dispositions and discourses,
along with the multiplicity of interpretation that is generated by these,
many a thinker has given up on the idea of an absolute truth. If truth is
only meaningful relative to a context or framework, we must ask: what is a
context? Either one accepts the truth of a given definition of context,
embracing once more the notion of an absolute truth, or one accepts that the
relation between truth and context is a relative one: truth is defined by
the framework that produced it, and the framework is that in which certain,
given truths hold. We believe this problem of the nature of truth, and
especially the popular tendency to congregate around an extreme
interpretation of the idea of relative truth, has undermined the believe of
rationality as a way to approach truth. Yet there are branches of human
activity that embrace the absolute notion of truth, simply because in some
cases, one cannot afford to succumb to relative truth. Take a court trial as
an example. Despite all various and often contradictory accounts of
witnesses, a judge or jury \textit{must} believe that some unique set of
events has happened which are to be judged as criminal or not by the court.
It is an implicit but nevertheless crucial assumption that lies at the basis
of our judicial system, that one single reality has given rise to all
various accounts, and not that every testimony has the ability to create its
own, independent reality. The exact sciences, by their very nature, were
also among those that have shown great reluctance to accept the notion of
relative truth. We can find a paradigmatic example of this struggle in the
wave--particle duality in the description\ of elementary particles or
quanta. Another area of research that faces similar problems, \textit{i.e.}
a plurality of descriptions of the same signal, is the field of signal
analysis. Therefore, it may not come as a great surprise that there are
certain characteristic features in the mathematical description of these two
theories that display similarities. Yet both fields are usually considered
only connected on a mathematical level, but to such extent that quite a few
of the more important definitions in signal analysis are borrowed from
earlier developments in quantum physics. We believe a very deep connection
can be identified in the way these theories relate the results of
observation to the state of the system which was observed. We will give
brief, mildly technical introductions to both fields of study and invoke a
metaphor from analytic geometry that, we believe, shows why the similarities
between the two are not coincidental, but emerged from the problem of
attributing truth to different, incompatible perspectives of the same
phenomenon. Finally we argue for the relevance and even necessity of
combining mutually incompatible perspectives in other fields of knowledge.

\section{The birth of quantum mechanics}

In 1925 a radically new theory of the fundamental constituents of matter and
energy was born. The final form the theory was to take, emerged from two
different developments, one due to Heisenberg, the other to Schr\"{o}dinger.
The first one started from Bohr's correspondence principle. In a nutshell
the principle says that although classical concepts, such as an electronic
orbit, could very well be meaningless concepts for elementary particles, the
classical level should somehow emerge as some limiting case. Heisenberg had
found that the mechanical equations that are usually written for the
positions and velocities of a physical system, are better expressed in terms
of the frequencies and amplitudes one measures in spectroscopic experiments
that form the main experimental source of data of the atomic realm. The
resulting theory was called matrix mechanics because the main quantities in
the classical theory of matter had to be replaced by matrices. The other
development took as a starting point the investigations of de Broglie's
conception of matter waves. de Broglie had postulated that with each
particle we can associate a wave that has a frequency that is related to its
mass. Schr\"{o}dinger's idea was to set up a wave equation for the matter
wave of an electron revolving around an atomic nucleus. In the beginning of
1926 he succeeded in deriving energy values for the states of a hydrogen
atom that coincided with the observed energy levels in spectroscopic
experiments. Later Schr\"{o}dinger showed that the two seemingly different
approaches are equivalent in that they derive the same values for the
operational quantities that they reproduce. When Schr\"{o}dinger realised
that the energy levels he had obtained for the hydrogen atom were simply the
frequencies of the stationary matter wave, he attempted to interpret the
wave as a matter wave. Matter would be spread out through space. This
smeared out matter would behave as a wave and the frequencies observed would
simply be a manifestation of the wave being confined around the nucleus.
Similar to a flute that produces a tone that has a wave length which
(possibly after multiplication by a small integer to account for the higher
harmonics) is equal to the length of the pipe in which it occurs, so would
an electron produce a light frequency in accordance with the length of its
orbit around the nucleus of the atom. However natural this assumption may
have seemed at the time, it became obvious the position was only tenable for
the most simple systems. The reason being that the wave does not travel in
ordinary space, but rather in the so-called configuration space, which has
more than three dimensions if the physical system consists of more than one
particle. The interpretation of the Schr\"{o}dinger waves that would stand
the test of time, was given by Born, who building on an idea of Bohr,
Kramers and Slater, proposed to interpret the wave as a probability wave. As
Heisenberg writes: \textquotedblleft something standing in the middle
between a reality and possibility\textquotedblright \cite{Heisenberg}. The
resulting theory gave surprising results in many respects. Classically, the
state of a system is characterised by its momentum and position. The space
of possible positions and momenta, is called the phase-space associated with
that system. All other quantities are simply a function of these two
variables, \textit{i.e.} a function of its location in phase-space. In the
new quantum mechanics, the observable quantities of a system, such as its
energy, or momentum, are still derivable from the state of the system, but
no longer by means of a simple function, but by an \textit{operator} acting
on the state of the system. The problem of finding the appropriate quantum
mechanical operator $\hat{f}$ for a given classical function $f$, is called
the quantization of that function $f$\footnote{This is not to be confused with the term \textit{quantization} that is
sometimes found in the literature on signal analysis, where it is meant to
denote a conversion of a datum into a zero or a one according to a
Neyman--Pearson criterion. See, for example, Ref. \cite{Helstrom},
p176.}. Quantization is by no means a trivial procedure. In fact, it has been
shown that no unique answer to the problem of quantization can be given in
general. Whereas clearly $f.g,$ $g.f$ and $(f.g+g.f)/2$ are the same
functions, the same expressions but with the functions $f$ and $g$ replaced
by their quantized counterparts $\hat{f}$ and $\hat{g}$, represent different
operators in general. Operators do not necessarily commute, \textit{i.e.} it
may very well turn out that 
\begin{equation*}
\hat{f}.\hat{g}\neq \hat{g}.\hat{f}
\end{equation*}%
Classically, it makes no difference to ask first what the position of a
system is and then its velocity or vice versa. But quantum mechanically, it
does.

\section{Whose point of view?}

We owe to Galilei the realization that quantities such as speed have a
relative character\footnote{It is part of physics folklore to attribute the principle of the relativity
of motion to Einstein. As a matter of fact, the notion was made precise by
Galilei almost 300 years before the advent of Einstein's theory of
relativity.}. The relevant question with regard to the motion of a body is
not \textquotedblleft What is the speed of a body?\textquotedblright , but
rather \textquotedblleft What is the speed of a body with respect to a
certain reference frame?\textquotedblright . Of course, the same goes for
quantum mechanics. However, there is an even deeper form of dependence on
the perspective of observation in quantum mechanics. To see how this works,
we will present an example from analytic geometry that provides a precise
analogy with quantum mechanics. If you never understood the problem of
non-commuting observables in quantum mechanics, but have no problem with
two-dimensional geometry, this analogy is for you.

\subsection{Description and frame of reference in analytic geometry}

The mathematical formula representing a hyperbole in the plane is presented
in standard elementary textbooks in the following form:%
\begin{equation}
x^{2}-y^{2}=1  \label{hyper1}
\end{equation}

\begin{figure}[tbh]
\begin{center}
\includegraphics[scale=0.5]{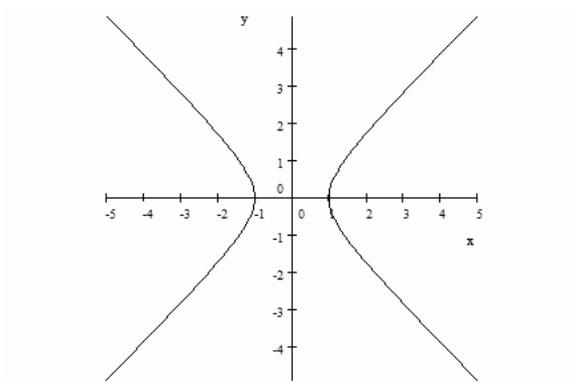}   
\end{center}
\caption{The graph corresponding to equation (1).}
\end{figure}

Next take a look at the following function 
\begin{equation}
\sqrt{2}xy=1  \label{hyper2}
\end{equation}

\begin{figure}[tbh]
\begin{center}
\includegraphics[scale=0.5]{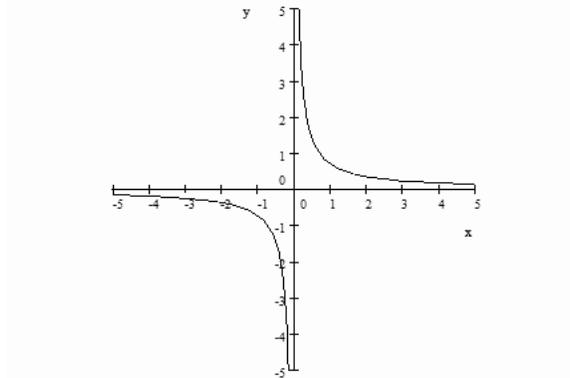}   
\end{center}
\caption{The graph corresponding to equation (2).}
\end{figure}
It is not obvious to determine at sight the relation between the two
prescriptions (\ref{hyper1}) and (\ref{hyper2}). However, when we plot the
graph it is rather obvious that (\ref{hyper2}) also represent a hyperbole.
Upon inspection, we see that graph 2 is like graph 1, but turned 45 degrees
anti-clockwise. The graphs make clear what the mathematical prescriptions do
not: that (\ref{hyper1}) and (\ref{hyper2}) represent the same geometrical
object but seen through different eyes. When we say it is the \textit{same}
geometrical object, we mean that one can be obtained from the other
employing only a combination of rotation, scaling and translation. This
corresponds to our experience in the real world. When an observer moves
through space, he may see an object at a different angle, a different size
and in a different location, but it will still be the same object. There is
a general mathematical method to see if two mathematical prescriptions of a
quadratic form such as (\ref{hyper1}) and (\ref{hyper2}) determine the same
geometrical object. First we have to associate a 2x2 matrix with the
mathematical prescription. To do so, we first write the mathematical
prescription in a canonical form%
\begin{equation}
a_{11}x^{2}+2a_{12}xy+a_{22}y^{2}=1  \label{quadratic equation}
\end{equation}%
The associated 2x2 matrix is a mathematical entity with 4 entries organized
as a square:%
\begin{equation}
A=%
\begin{pmatrix}
a_{11} & a_{12} \\ 
a_{21} & a_{22}%
\end{pmatrix}
\label{matrix}
\end{equation}%
So $a_{11}$ is the coefficient of $x^{2},$ $a_{22}$ is the coefficient of $%
y^{2}$, $a_{12}$ is half the coefficient of $xy$ and $a_{21}$ half the
coefficient of $xy.$ This means that $a_{12}=a_{21}.$ In general (that is,
also for more than two free variables), the matrix of a quadratic form is 
\textit{symmetric}. This means that if we interchange the rows and columns
of $A,$ (this is the so-called transpose $A^{T}$ of that matrix) we get the
same matrix: 
\begin{equation}
A^{T}=A  \label{symmetric}
\end{equation}%
Next we introduce a vector $\mathbf{x,}$ which is a shorthand for the two
original variables $x$ and $y,$and its transpose $\mathbf{x}^{T}$: 
\begin{equation}
\mathbf{x}=%
\begin{pmatrix}
x \\ 
y%
\end{pmatrix}%
\text{ and }\mathbf{x}^{T}=%
\begin{pmatrix}
x & y%
\end{pmatrix}
\label{vectors}
\end{equation}%
We are now in a position to write any (central) conic sections, \textit{i.e.}
the circle, the ellipse, the parabola and the hyperbole\footnote{%
The same equation but with three dimensional vectors, and hence with $A$ a
3x3 matrix, describes an even greater plurality of geometrical objects
called \textit{quadrics} for obvious reasons.} in the quadratic form: 
\begin{equation}
\mathbf{x}^{T}A\mathbf{x}=1  \label{quadratic form}
\end{equation}%
For example, when employing the rules for matrix multiplication (which we
will not give here), we get
\begin{equation}
\mathbf{x}^{T}A\mathbf{x}=%
\begin{pmatrix}
x & y%
\end{pmatrix}%
\begin{pmatrix}
1 & 0 \\ 
0 & -1%
\end{pmatrix}%
\begin{pmatrix}
x \\ 
y%
\end{pmatrix}%
=x1x+y(-1)y=1  \label{matrix 1}
\end{equation}%
which clearly represents (\ref{hyper1}), and
\begin{equation}
\begin{pmatrix}
x & y%
\end{pmatrix}%
\begin{pmatrix}
0 & \sqrt{2}/2 \\ 
\sqrt{2}/2 & 0%
\end{pmatrix}%
\begin{pmatrix}
x \\ 
y%
\end{pmatrix}%
=x(\frac{\sqrt{2}}{2})y+x(\frac{\sqrt{2}}{2})y=1  \label{matrix 2}
\end{equation}%
which represents (\ref{hyper2}). Next we search for a way to identify the
shape of the solutions independent of the actual form of the $x$'s and $y$%
's. For this we diagonalize, that is transform it in such a way that only
the entries $a_{11}$ and $a_{22}$ are different from zero. Of course, we are
only allowed transformations of the matrix that leave the original
geometrical object it represented intact. For an arbitrary matrix this is
not possible, but for a symmetric 2x2 matrix, as is the case here, we can
always bring it into a diagonal form by only applying a single rotation. If
we applied the method of diagonalization (which we will not present here),
we would find the diagonal form of the 2x2 matrix $A$ in (\ref{matrix 2}) is
precisely the 2x2 matrix in (\ref{matrix 1}). In this new reference frame,
obtained from the old by rotation, we can easily recognise the kind of
geometrical object the quadratic form presents, because it is in its
standard form. There is important terminology to remember here. The axes of
the new frame of reference correspond to the \textit{eigenvectors} of the
matrix $A$. What is an eigenvector?

For a moment, let us forget about the original functions we are studying and
just look at the following matrix equation:%
\begin{equation}
A\mathbf{x}=\mathbf{x}^{\prime }  \label{lin transformation}
\end{equation}%
This equation tells us every vector $\mathbf{x}=(x,y)$ will, upon
multiplication with the matrix $A,$ transform into a new vector $\mathbf{x}%
^{\prime }$. Of course it could be the case that for a certain special
vector $\mathbf{x}_{0}$ this new vector $\mathbf{x}_{0}^{\prime }$ is just a
multiple of the old vector $\mathbf{x}_{0}$:%
\begin{equation}
A\mathbf{x}_{0}\mathbf{=\mathbf{x}_{0}^{\prime }=\lambda }_{0}\mathbf{x}_{0}
\label{eigenequation}
\end{equation}%
If this is the case, we say $\mathbf{x}_{0}$ is an \textit{%
eigenvector} of the matrix $A$ and the (real or complex) number $\mathbf{%
\lambda }_{0}$ is its associated \textit{eigenvalue}. For example, the
eigenvalues of the matrix in (\ref{matrix 1}) are $1$ and $-1$. Armed with
this terminology we can make a summary of our observations.

\textbf{Summary}: \textit{The quadratic form (\ref{quadratic form}) obtains
its most simple (diagonal) form when described in a frame of reference whose
axes are the vectors (\ref{eigenequation}) that are left invariant (up to
multiplication by a number) by the matrix (\ref{matrix}) associated with the
quadratic form. When presented in this form, the diagonal elements of the
matrix }$A$\textit{\ are called the eigenvalues\ of }$A$\textit{. The
geometrical content of the quadratic form (\ref{quadratic form}) can be
determined directly from the eigenvalues\ of its associated matrix.}

If we have this diagonalized form, we can easily recognise whether the
geometrical object is a parabola or a hyperbole or an ellipse, because the
diagonal form is just the standard function we learned for these objects in
high school. If the eigenvalues of the matrix in (\ref{matrix 1}) are $1$
and $-1$ (as is the case here), this means the geometrical object is a
hyperbole. If both had been $+1$, it would have been a circle. If they had
been two different positive numbers, it would have been an ellipse. When one
of them is zero, but not both, we are dealing with a parabola.

Of course, the same results can be obtained using only elementary algebraic
calculations, but the method presented here works for an arbitrary number of
dimensions, not just for two, as long as we are only presenting a quadratic
form. It is important to note that it works for vectors and matrices with
complex entries too, but the condition (\ref{symmetric}) of symmetry is to
be replaced with the corresponding complex notion, that of a self-adjoint
matrix. Think of a self-adjoint matrix as simply a matrix that can be
brought into diagonal form and for which the eigenvalues are all real\footnote{Technically, a self-adjoint matrix is one which is equal to its adjoint
matrix. The adjoint of a matrix is defined as the complex conjugate of its
transpose. This coincides with the definition given in the text.}. It turns
out that an observable quantity in quantum mechanics \textit{is}
represented by a self-adjoint operator. Indeed, the same technique can be used
to describe systems with a finite number of outcomes
quantum mechanically, and the analogy is quite precise.

\subsection{Incompatibility in analytic geometry}

Suppose we have an equation such as%
\begin{equation}
-\sqrt{2}x^{3}y+\sqrt{2}xy^{3}+\sqrt{2}xy+x^{2}-y^{2}=1  \label{big formula}
\end{equation}%
what geometrical object would this represent? After a simple calculation, we
find that it is equivalent to%
\begin{equation}
(x^{2}-y^{2}-1)(\sqrt{2}xy-1)=0  \label{hyper 1+2}
\end{equation}

What is the solution of this equation? Well, it is of the general form $AB=0$
with $A=x^{2}-y^{2}-1$ and $B=\sqrt{2}xy-1.$ This equation is satisfied when
either $A$ is zero, or $B$ is zero. So the equation is obviously satisfied
when either (\ref{hyper1}) or (\ref{hyper2}) is satisfied. The complete
solution of (\ref{hyper 1+2}) is then the set of points $(x,y)$ that belongs
to the first or the second hyperbole, and the graph of the horrible looking (%
\ref{hyper 1+2}) is nothing but two graphs of (\ref{hyper1}) and (\ref%
{hyper2}) together. The relevant observation here, is that the basis that
makes (\ref{hyper1}) diagonal (and hence takes the formula for the hyperbole
in its standard form) does not make (\ref{hyper2}) diagonal, simply because
the eigenvectors of the two are different. We can transform the form (\ref%
{big formula}) such that either (\ref{hyper1}) is in standard form, or (\ref%
{hyper2}), but never the two simultaneously. We could say the two standard
forms are incompatible in the sense that no single perspective can yield
both simple forms at the same time. A similar situation occurs in the
description of quantum systems.

\subsection{Description and complementarity in quantum mechanics}

Suppose we have a system that we seek to describe by quantum mechanics. The
theory tells us the state of a physical system is represented by a vector,
say $\psi $, living in an abstract Hilbert space $H$\footnote{A Hilbert space is a generalisation of the well-known vector space, in that
it allows for infinite dimensions and is equipped with an inner product.}.
Another fundamental postulate of quantum mechanics is that any quantity that
we want to infer experimentally from that state $\psi $, whether it be its
energy, its position or its momentum, is represented by a self-adjoint
operator $A$. Quantum mechanics dictates furthermore that the outcome of the
measurement will be one of the eigenfunctions of the operator,\footnote{%
If the Hilbert space of the system is infinite dimensional, not all
operators have eigenfunctions which are vectors in $H$ (eigenvectors).
None-the-less, the interpretation of a measurement of $A$ is the same.} and
that the \textit{probability} of obtaining that result is the square of the
norm of the \textit{coefficient} that goes with the eigenvector which
corresponds to the outcome. Furthermore, if the matrix $A$ represents the
quantity we are interested in, and we want to know the expectation value $%
\langle A\rangle $ (a number representing the average physical value of $A$) when the system is in the
state $\psi $, we have 
\begin{equation}
\langle \psi |A|\psi \rangle =\langle A\rangle  \label{expectation}
\end{equation}%
We briefly indicate the meaning of the notation in (\ref{expectation}). The
quantity $\langle A\rangle $ on the right-hand side of equation (\ref%
{expectation}) is just a number. We can always rescale the matrix so that
the left-hand side of the equation equals one without altering the essential
geometrical content \footnote{Of course, this changes the expectation value itself, but this is of no
consequence. One can also change the value of any quantity by scaling the
appropriate dimensions, without in any way changing the physics of the
phenomenon. Our rescaling only serves the purpose of demonstrating the
essential equivalence between (\ref{expectation}) and (\ref{quadratic form}).}. The so-called bra vector $\langle \psi |$ in (\ref%
{expectation}) is the dual of its corresponding ket vector $|\psi \rangle $,
just in the same way as $\mathbf{x}^{T}$ in (\ref{quadratic form})
represents the dual of $\mathbf{x}$ for finite spaces. These are the
fundamental postulates of simple, non-composite, non-relativistic quantum
systems and to understand the geometrical meaning of the words in these
postulates, we only need the analogy with analytic geometry. Indeed, Eq.~(\ref{expectation}) may look cumbersome to the reader who is unfamiliar with
the \textit{bra-ket} notation, but for observation of systems with
only a finite number of outcomes, it is formally equivalent to the quadratic
form (\ref{quadratic form}):%
\begin{equation*}
\mathbf{x}^{T}A\mathbf{x}=1
\end{equation*}%
According to the former section on analytic geometry, we already know how to
find the eigenvectors and corresponding eigenvalues: we simply have to
diagonalise the self-adjoint matrix that corresponds to that observable
quantity. Just as in the case of analytic geometry, the matrix $P$
corresponding to a given operator, say its momentum $\hat{p}$, yields its
\textquotedblleft secrets\textquotedblright\ easily in a description that
fixes the axes of reference to the eigenvectors of that same matrix: in its
diagonal form, the probability of an outcome is simply the square of the
corresponding diagonal element. Now we ask whether the same is true for the
position of the system. Well of course, it is an axiom of quantum mechanics!\footnote{%
Mathematically speaking, we are making an unwarranted simplification. The
operators coresponding to $Q$ and $P$ live in an infinite dimensional space,
for which we have to use the spectral theorem, rather than speak about
eigenvalues and eigenvectors. The reader is asked to keep this in mind and
understand that we are really talking about the infinite dimensional
equivalents of the terms we employ here.}. Simply calculate the eigenvectors
of the matrix $Q$ that corresponds to the position operator $\hat{q}$ of a
system, and draw your frame of reference there. Look again at what you have
got, and you will only find diagonal entries from which you can immediately
read the probability of obtaining a certain result. Now we may wonder: is
the new frame of reference that diagonalizes $Q$, also a frame of reference
that diagonalizes $P$? The answer depends on the relation between $P$ and $Q$%
. It is a mathematical theorem that if and only if\ $P$ and $Q$ (or their
associated operators $\hat{p}$ and $\hat{q}$ ) \textit{commute}, there
exists a common frame of reference in which both are diagonal. But $P$ and $%
Q $ do \textit{not} commute in quantum mechanics, and hence there is a
problem with the simultaneous extraction of position and momentum
information for a quantum mechanical system. Indeed, in quantum mechanics,
we have the commutator relation:%
\begin{equation}
\hat{p}\text{ }\hat{q}-\hat{q}\text{ }\hat{p}=\frac{ih}{2\pi }\hat{I}
\label{commutator qm}
\end{equation}%
in which $h$ is the famous constant of Planck and $\hat{I}$ the identity
operator which is clearly different from zero \footnote{It is known since the advent of quantum mechanics that no finite
matrices $P$ and $Q$ can satisfy (\ref{commutator qm}), once more indicating
the necessity of upgrading the traditional vector space of analytic geometry
and finitary quantum mechanics to the Hilbert space.}. So we know from this
equation (whatever the value at the right hand side was) that there does not
exist a common frame of reference such that both $P$ and $Q$ are diagonal.
Bohr introduced the notion of complementarity to describe this phenomenon. $%
Q $ and $P$ are like the inside and outside view of a house: both yield
valuable information as to what a house is, but the better one can see the
inside, the more one is neglecting the outside. Bohr coined the phrase that $%
P$ and $Q$ are \textit{complementary} descriptions of the same system. The
simultaneous extraction of $P$ and $Q$ information is hence limited by the
uncertainty principle of Heisenberg, to which we will come back later. Many
of the early gedanken experiments in the early history of quantum mechanics,
and from the historic Solvay conference in 1927 in particular, can serve
(and indeed were meant to serve) as an illustration of this principle \cite{Wheeler and Zurek}.

\section{Signal analysis}

There is another field of scientific inquiry that has had to learn to deal
with incompatible perspectives of the nature of the phenomenon; the field of
signal analysis. The field of signal analysis is concerned with the
extraction of information from a time-varying quantity which is called the
signal. The signal can originate from very different sources. It can be
electromagnetic, acoustic, optical or mechanical. It can contain various
quantities and qualities of information. Needless to say, how to extract
information depends on the way you look at things. Let us take an example.
Suppose a sound technician is responsible for the music played by a live
band at a party with a strict beginning and end in time and a strict limit
to the amount of decibels that may be produced so that no neighbors are
disturbed by the party. To make sure he starts and ends at the right moment,
he checks, and if necessary adjusts his watch beforehand. At the moment the
band starts, he lets the volume rise to an audible level that is still under
the prescribed decibel limit and stops right before the announced end of the
party. Take the music coming out of the speakers as the signal. The
mathematical representation of this music as a time-varying quantity will
allow him to do all that he needs. Given just this time-varying quantity, he
could even have been in a remote place where he cannot hear the music and
still perform the task he was given with a remote control. He merely needs
to check that the signal starts after the prescribed time, stops before the
prescribed ending, and make sure it does not exceed the given maximum
volume. Suppose now however that the live music produces a feedback effect.
The volume grows and an insistent, painful whistling tone is heard. If the
sound technician is there, he will correct the problem, but if he is in a
remote location with his remote control, he will simply lower the volume
somewhat and assume everything is alright. The people attending the party
will probably judge differently. The time varying quantity does not easily
allow for the reading of frequency related information. If however, the
sound technician made use of a spectral analyzer, he would have been able to
correct even that problem remotely. The spectral analyzer performs a
transformation of the signal. Rather than displaying the intensity of the
signal in time, it yields the intensity of the various frequency bands heard
within a given time interval. The feedback will be visible as a rapid and
lasting rise of a small frequency band. Technically, the transform involved
is called a short-time Fourier transform of the signal. It turns out that
the operator corresponding to the Fourier transform does not commute with
the operator that yields the intensity at every instant. So we see that the
situation is quite similar to quantum mechanics from this point of view. In
both situations, we are looking for an appropriate perspective to describe
or extract the information, by transforming the system to a frame of
reference that is constructed on the basis of the eigenvectors of the
quantity that we seek to describe\footnote{Again we reach the limit of the validity of our analogy. We should have said
that it is impossible to construct spectral families for the operators
simultaneously. In fact, to do the job properly, we need to decompose the
operators in terms of so-called positive operator valued measures rather
than positive valued measures (spectral families). For the interested
reader, we refer to \cite{Schroeck}.}.

\subsection{Quantum probability and signal analysis}

The introduction of many of the definitions and methods used in signal
analysis originated from similar developments in quantum mechanics. In a
sense this is a curious historical fact, as it was originally accepted that
quantum physics presents a fundamentally different theory than classical
physics and the physics of signals is essentially classical. As physicists
like to point out, quantum mechanics is inherently indeterministic, and
signals are essentially deterministic. Whereas quantum physics describes the
fundamental interactions in nature and is governed by Planck's constant,
signals are continuous in nature and no constant governs their behavior. Yet
in the literature on signal analysis, see \cite{Flandrin} and \cite{Mallat}, we find the names of famous quantum physicists like Weyl (ordered
operator algebra), Wigner (quasi-distribution), Heisenberg (uncertainty
boxes) and Cohen (bilinear class). In 1946 Gabor, the inventor of the
hologram and the atomic representations of signals, introduced the
Heisenberg inequality in signal analysis \cite{Gabor}. Only two years later,
Ville,\cite{Ville} was to introduce a time--frequency distribution as a
signal analytic tool that is formally equivalent to Wigner's distribution in
quantum mechanics \cite{Wigner}. The similarities between the two is of such
extent, that apparently some concepts were invented independently in both
subjects without reference to its parallel in the other subject. Although
Ville constantly evokes language borrowed from quantum mechanics, the Ville
distribution itself was originally given without reference to Wigner and the
same occurred with the Rihaczek distribution,\cite{Rihaczek} that was
historically preceded by the Margenau and Hill distribution in 1961 \cite{Flandrin}. Sometimes researchers in quantum mechanics would cross the border
to publish results useful to the signal analyst. We mention Cohen's work, on
bilinear and completely positive phase space distributions and the squeezed
states of Caves,\cite{Caves} and Man\'{}ko,\cite{Manko} that have found
application in the construction of signals that are used in radar
technology. To give the reader a taste of the extent of the overlap between
the two fields, we sum up the main similarities\footnote{The reader who is not familiar with either subject is advised not to take
the terminology too seriously and just regard this list as a witness to the
similarities.}. Both signals and quantum states are identified with vectors
in a Hilbert space. Both see as operational quantities the modulus squared
of these elements, which in signal analysis is called the energy density of
the signal and in quantum mechanics the probability density of the entity.
Both utilize the Fourier transform as the relationship between two of the
most important operational quantities, time and frequency in signal analysis
and position and momentum in quantum mechanics. As a consequence, both use
the Heisenberg inequality and frequently employ non-commuting operators for
observable quantities. Both fields employ the Wigner quasi-distribution as a
phase space quasi-density.

\subsection{The Fourier paradigm}

Basic detection and error estimates can be given just from examining the
major repetitive components of a signal. If major components in the signal
tend to have a repetitive nature, one can average the signal over a large
number of cycles. This technique is called time synchronous averaging and
diminishes the noise content of the signal. Another very popular technique
to acquire primitive frequency-related information of a signal $\phi (t)$,
is to take the auto-correlation function, which is the average of the
product $\phi (t)\phi (\tau +t).$ However, if the signal is too complex,
these methods have their limitations in providing frequency related
information. The cure, which marked the birth of modern mathematical signal
analysis, dates back to Joseph Fourier's (1768--1830) investigations into
the properties of heat transfer in the early 1800's \cite{Fourier}. Fourier
conjectured that an arbitrary periodic function, even with discontinuities,
could be expressed by an infinite sum of pure harmonic terms. The idea was
ridiculed among many great scholars, but a large portion of mathematical
analysis in the 19th and 20th centuries was dedicated to making the
statements of Fourier precise and the Fourier transformation eventually
became a tool of utmost importance in signal analysis, quantum physics and
the theory of partial differential equations. In words, the Fourier theorem
states that the sum of an infinite number of appropriate trigonometric
functions, can be made to converge to any periodic function. Take a signal
with period $2L$%
\begin{equation*}
\phi (t)=\phi (t+n2L),n=1,2,3,\ldots 
\end{equation*}%
Calculate the following quantities with $\omega = \frac{\pi }{L}$:
\begin{eqnarray*}
a_{0} &=&\frac{1}{L}\int_{-L}^{L}\phi (t)dt \\
a_{n} &=&\frac{1}{L}\int_{-L}^{L}\phi (t)\cos (n\omega t)dt \\
b_{n} &=&\frac{1}{L}\int_{-L}^{L}\phi (t)\sin (n\omega t)dt
\end{eqnarray*}%
Knowledge of the signal $\phi (t)$ obviously implies knowledge of the
numbers $a_{n}$ and $b_{n},$ provided the integrals exist. But Fourier's
theorem states the far less obvious fact that, under mild assumptions, the converse also holds
\begin{equation*}
\phi (t)=a_{0}/2+\sum_{n=1}^{+\infty }a_{n}\cos (n\omega t)+b_{n}\sin
(n\omega t)
\end{equation*}%
Knowledge of the numbers $a_{n}$ and $b_{n}$ implies complete knowledge of $%
\phi (t).$ The Fourier expansion is such a valuable tool because of its
ability to decompose any periodic function (\textit{e.g.} machine
vibrations, sounds, heat cycles,\ldots ) by means of sines and cosines. As
the reader may now appreciate, one can regard the sines and cosines as the
eigenfunctions of the frequency operator. They form the orthonormal basis
functions, the axes of the new frame of reference, and the coefficients of
these orthonormal basis functions represent the contribution of the sine and
cosine components in the signal. This allows the signal to be analyzed in
terms of its frequency components. Rather than studying signals in their
time domain, one could now study the characteristic frequencies of the
phenomenon. A serious draw-back of the method is that only periodic signals
can be represented. The Fourier paradigm was extended to cover non-periodic
functions by means of the so called Fourier transform. To include
non-periodic functions, we take the limiting case of the Fourier series when
the period of the function goes to infinity. When we do this, we obtain the
following formulas: 
\begin{eqnarray}
\Phi (\nu ) &=&\int_{-\infty }^{+\infty }\phi (t)\exp (-2\pi i\nu t)dt
\label{fourier pair} \\
\phi (t) &=&\int_{-\infty }^{+\infty }\Phi (\nu )\exp (2\pi i\nu t)d\nu  
\notag
\end{eqnarray}%
We call $\Phi (\nu )$ the Fourier transform of $\phi (t),$ often written as $%
\mathcal{F}[\phi (t)]=$ $\Phi (\nu ),$ and $\phi (t)$ is the inverse Fourier
transform of $\Phi (\nu )$ written as $\mathcal{F}^{-1}[\Phi (\nu )]=\phi (t)
$. The couple of functions $\{\phi (t),$ $\Phi (\nu )\}$ is called a Fourier
pair. The drawback of this representation, as one can judge from the
boundaries of the integral, is that the signal has to be known at all
possible times, something that is never the case in applications. Of course,
in an idealized situation, such as a quasi-monochromatic or a steady state
description, the method yields true properties of the signal in question.
But transitional regimes of signals are simply beyond the scope of this
method. Even with this limitation, the importance of the consequences of 
(\ref{fourier pair}) can hardly be overestimated. It was a source of
inspiration to most advanced methods of transformation that eventually led
to the field of wavelets, which have found wide spread use in almost all
practical branches of contemporary applied science \cite{Mallat}. Let us
therefore have a brief look at some of the properties of the Fourier pair (\ref{fourier pair}).

\subsection{Properties of Fourier pairs}

We call the region were a function does not vanish (\textit{i.e.} is not
equal to zero), the \textit{support} of that function. If the support of $%
\phi (t)$ is of finite width, then the width of the support of $\Phi (\nu )$
is infinite and \textit{vice versa}. But even with unbounded support (%
\textit{i.e.}, when the support has infinite extension), a reciprocal
relationship exists between $\phi (t)$ and $\Phi (\nu ).$ To see this,
assume the signal is normalized ($\int_{-\infty }^{+\infty }|\phi
(t)|^{2}dt=1$) and that the center of gravity of both $\phi (t)$ and $\Phi
(\nu )$ vanishes:%
\begin{equation*}
\int_{-\infty }^{+\infty }t|\phi (t)|^{2}dt=\int_{-\infty }^{+\infty }\nu
|\Phi (\nu )|^{2}d\nu =0
\end{equation*}%
This can always be realized by an appropriate choice of the axes. We define
the following quantities: 
\begin{eqnarray*}
\Delta t &=&\int_{-\infty }^{+\infty }t^{2}|\phi (t)|^{2}dt \\
\Delta \nu &=&\int_{-\infty }^{+\infty }\nu ^{2}|\Phi (\nu )|^{2}dt
\end{eqnarray*}

If $\phi (t)$ decays fast enough such that $t|\phi (t)|^{2}$ vanishes at infinity, the
following relation follows from the Cauchy--Schwartz inequality:%
\begin{equation*}
\Delta t.\Delta \nu \geq \frac{1}{4\pi }
\end{equation*}%
This relation is known in the literature on signal-analysis as the
Heisenberg--Gabor uncertainty principle. The name is due to the fact that
Gabor, an electrical engineer working on communication theory, discovered
the relevance of this relationship to signal-analysis in 1946, about two
decades after Heisenberg --- one of the founding fathers of quantum physics
--- formulated the principle on physical grounds \cite{Gabor}. Weyl showed it
to be a mathematical consequence of the fact that the corresponding
operators are a Fourier pair \cite{Weyl}. And we see that indeed, the
frequency and temporal descriptions of a signal yield complementary views of
its information content. It should therefore come as no surprise that their
respective operators do not commute. Instead of $\hat{t}$ $\hat{\nu}=\hat{\nu%
}$ $\hat{t},$ we have: 
\begin{equation*}
\hat{\nu}\text{ }\hat{t} - \hat{t}\text{ }\hat{\nu} =\frac{i}{2\pi }\hat{I}
\end{equation*}%
Similar equations exist for the time and scale operators that are found in
the literature on wavelets. This is to be compared with the equivalent
relation for position and momentum in quantum mechanics (\ref{commutator qm}%
):%
\begin{equation*}
\hat{q}\text{ }\hat{p}-\hat{p}\text{ }\hat{q}=\frac{ih}{2\pi }\hat{I}
\end{equation*}%
where $h$ denotes Planck's constant. This latter relation implies the
well-known Heisenberg uncertainty relation:%
\begin{equation*}
\Delta q.\Delta p\geq \frac{h}{4\pi }
\end{equation*}%
We see that, apart from Planck's constant, the commutator and uncertainty
relations have the same mathematical content when we apply the following
substitution:%
\begin{eqnarray*}
\hat{p} &\rightsquigarrow &\hat{\nu} \\
\hat{q} &\rightsquigarrow &\hat{t} \\
h &\rightsquigarrow &1
\end{eqnarray*}%
With this basic vocabulary, it is easy to extend the list to include many
more examples of corresponding quantities. We refer the interested reader to Ref.~\cite{Cohen}, p. 197.

\section{Concluding remarks: The necessity of combining mutually
incompatible perspectives.}

We have seen that there are striking mathematical similarities between
quantum mechanics and signal analysis. Historically, these mathematical
analogies are the roots of the ongoing cross-fertilization between the two
fields. Moreover, when we encounter a discussion in the literature that
attempts to answer \textit{why} this cross-fertilization has taken place,
the similarity between the mathematics is invariably seen as the main
source. There is no doubt that this analysis of the situation is correct,
but it leaves the question \textit{why} the similarities are there in the
first place unanswered\footnote{It is interesting to note that detailed similarities can be explained as a
result of the irreducible representations of the Poincar\'{e} group. These
representations are the same for both the massive and zero mass cases, so
they apply to photons, which are often the carriers of signals and massive
quanta alike. The Lie algebra is the same for both and that gives rise to
the uncertainty relations. For more details, see \cite{Schroeck}.}. Of
course, the analogies could be purely coincidental. For example, both fields
emerged from studies in the physics of heat. Quantum physics originated from
the study of black body radiation, and signal analysis was born with
Fourier's study of heat processes. We believe this is indeed coincidental,
as neither quantum physics, nor signal analysis deals primarily with the
problems of heat transfer or radiation. On the other hand we believe the
nature and origin of the similarities we have described is not coincidental,
and we have given a metaphor from analytic geometry to support that thesis.
The origin of the analogy that we identified from our metaphor is that, even
though the state of a system (or signal) can be fully known, the extraction
of measurable quantities (relevant information of signal) from a state
requires taking a point of view, and not all points of view are compatible
with one another. So we conjecture that the reason for the mathematical
parallels is that both theories inherently contain a theory of observation.
Complementarity arises in both fields because in some parts of reality there
are mutually incompatible, but equally valid, perspectives of the same
object. One cannot see the inside and outside of a house at the same time. A
natural question to consider is then: When do we know when we have a
sufficient number of views so that we are certain that we have seen
\textquotedblleft the full house\textquotedblright ? The metaphor from
analytic geometry regards \textquotedblleft perspective\textquotedblright\
as the equivalent of \textquotedblleft frame of reference\textquotedblright
. We have elaborated that the frame constructed from the eigenvectors of the
operator that represents the kind of information we are interested in yields
the most simple description. So our question can now be translated as
follows: What set of operators do we need to make sure that we have seen 
every aspect of the house? The most direct way of
tackling that question, relies on Prugove\v{c}ki's concept of \textit{informational completeness }within quantum mechanics \cite{Prugovecki}. A set of operators is called informationally complete if there
is only one state compatible with the probabilities obtained from it. This
means that for an informationally complete set of measurements, the state
can be uniquely derived from the expectation values of these measurements.
Moreover, the expectation value of \textit{any }operator (\textit{i.e.} any
point of view) can be calculated as a simple average of functions of the
outcome probabilities of the informationally complete set.

Now one can raise the question whether the whole phenomenon of
complementarity cannot be avoided. If we take a sufficient number of
compatible perspectives (\textit{i.e.}, commuting operators) can we obtain
an informationally complete set? The answer is \textit{no}. It was shown by
Bush and Lahti \cite{BL}, that no set of commuting observables is
informationally complete! Complementarity is here to stay. This was
intuitively accepted by most researchers in both fields, even though the
result of Bush and Lahti is not widely known among quantum physicists, and
largely unknown among signal analysts. Nevertheless, it is common knowledge
among researchers in the field of theoretical signal analysis, that
analyzing a signal in terms of the distribution in the non-commuting
observables that are relevant, provides a much more revealing picture in
many cases than, for example, the spectrogram or the signal together with
its Fourier transform. One can show the spectrogram (and most other
representations) to be informationally incomplete and the Wigner--Ville
distribution to be informationally complete \cite{Schroeck}.

If our analysis makes sense, we should find that if one restricts the
observations in quantum physics to the case of a single observable, that is
to the results of essentially one experimental set-up, the results of the
quantum probabilistic scheme can be reproduced by a classical probability
model. It turns out that this is true. Hence the power of quantum
statistical methods over classical probabilistic analysis is perhaps most
obvious when dealing with expectation values and correlations of observables
that do not commute. It is in this respect that quantum-like
representations, be it a phase space representation or a density matrix,
provide us with a statistically complete language for the incorporation of
data from disjoint or incompatible measurements into a single state of the
entity under observation. As a consequence, the conventional paradigm that
quantum probability pertains to the micro-world and classical
(Kolmogorovian) probability pertains to the macro-world is no longer tenable
from this point of view. This is not because micro-physical principles of
the entities under observation are at work, but because the relation between
the possible ways in which we look at these phenomena is structurally
isomorphic. But mutually incompatible perspectives are to be found in many
areas that are of scientific interest. What about the social implications of
interactions between cultures with incompatible value systems? What about
the well-known graphical figures that yield one gestalt or another (but
never both at the same time) depending on how they are looked at? A sentence
may be deemed innocent within one context, but disturbing in another. A
single note can make one chord cheerful and another chord sad. One cannot
hope that investigations in these areas will yield precise quantitative
statements any time soon in the same way as we have in engineering or
physics. Nevertheless, the lesson that it is only together that these
incompatible perspectives give an informationally complete picture may well
be still valid.

\section*{Acknowledgments}

Part of the research for this article took place in the framework of the
FWO-research community: \textquotedblleft The construction of integrating
worldviews\textquotedblright. The authors acknowledge financial support from FWO research project G.0362.03N.

\end{document}